\documentclass[twocolumn,aps,prb]{revtex4}
\usepackage{bm}  % bold math
\usepackage{graphicx}
\usepackage{amssymb}
\usepackage{color}
\usepackage{amscd}
\usepackage{epsfig}
\def\beq{\begin{equation}}
\def\eeq{\end{equation}}
\def\beqa{\begin{eqnarray}}
\def\eeqa{\end{eqnarray}}

\newcommand{\bfr}{{\bf r}}
\newcommand{\bfv}{{\bf v}}

\newcommand{\bfF}{{\bf F}}

\newcommand{\bfp}{{\bf p}}
\newcommand{\bnabla}{{\bm \nabla}}
\newcommand{\bxi}{{\bm \xi}}
\newcommand{\bomega}{{\bm \omega}}

\newcommand{\bsigma}{{\bm \sigma}}

\newcommand{\nhat}{\hat{n}}
\newcommand{\nvec}{{\bf \hat{n}}}
\newcommand{\uhat}{\hat{u}}
\newcommand{\uvec}{{\bf \hat{u}}}

\begin{document}
\title{Rheology of Active Filament Solutions}
\author{T. B. Liverpool$^{1,3}$}
\author{M. C. Marchetti$^2$}
\affiliation{$^1$Department of Applied
Mathematics, University of Leeds, Woodhouse Lane, Leeds LS2 9JT,
UK}
\affiliation{$^2$Physics Department, Syracuse University,
Syracuse, NY 13244, USA}
\affiliation{$^3$Kavli Institute for Theoretical Physics, UCSB, Santa Barbara, CA 93106, USA}
\date{\today}

\begin{abstract}
  We study the viscoelasticity of an active solution
 of polar biofilaments and motor proteins.
  Using a molecular model, we derive the constitutive equations for the stress
  tensor in the isotropic phase and in phases with
  liquid crystalline order. The stress relaxation
  in the various phases 
  is discussed. Contractile activity is responsible for a
  spectacular difference in the viscoelastic properties on opposite
  sides of the order-disorder transition.
\end{abstract}
\pacs{87.16.-b,82.70.Gg,05.65.+b}
\maketitle

Soft active systems are exciting examples of a new type of
condensed matter where stored energy is continuously transformed
into mechanical work at microscopic length scales. A realization
of this are polar filaments interacting with associated molecular
motors in the cell cytoskeleton \cite{Alberts,Howard}. These
systems are characterized by a variety of dynamic and stationary
states which the cell accesses as part of its cycle
\cite{takiguchi,nedelec97,surrey01}. Recent experimental and
theoretical studies of the dynamics of solutions of active
filaments have focused on the formation of both homogeneous and
inhomogeneous
%%isotropic and polarized
states with spatial structures, such as bundles, vortices or
asters
~\cite{nedelec97,surrey01,nakazawa96,Kruse00,TBLMCM03,Lee01,ramaswamy02,Kruse04,Sankararaman04,Aranson05,Voituriez}.

In this letter we study the effect of motor activity on the
rheological properties of active solutions under an externally imposed
stress. Understanding the viscoelasticity of cells, and cellular
extracts in-vitro, is a very important problem currently receiving a
lot of experimental attention
\cite{sackmann96,kas02,Shin2004,Claessen2006,Danuser2006}. From a
theoretical point of view describing the mechanics of the cytoskeleton
in its full complexity remains very challenging. As a first step in
this direction we focus here on using methods from polymer physics to
understand the effect of motor activity on the viscoelasticity of a
dilute solution of long stiff biopolymers. A phenomenological
description of the rheology of isotropic suspensions of active
particles near the isotropic-nematic transition was proposed recently
by Hatwalne et al.  \cite{hatwalne04}. The present work provides a
%%particular microscopic realisation of 
microscopic basis for 
their results and a general
framework for analyzing the viscoelastic behavior of active
solutions in both isotropic and ordered states.

\begin{figure}
\center \resizebox{0.45\textwidth}{!}{%
  \includegraphics{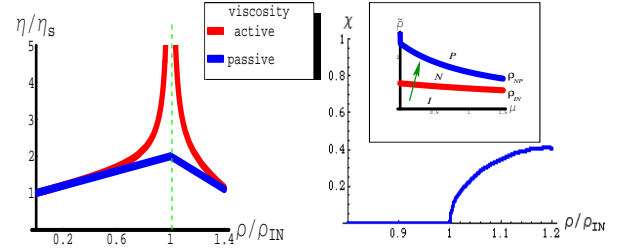}
%%%%
}

\caption{
  (a) The steady-state shear viscosity on both sides of the IN
  transition for passive and active nematics.(b) The 'motility'
  parameter, $\chi$ (see Eq.~(\ref{new_constit})) relating the magnitude of the active component of
  shear stress to the 'activity' $\mu$ ( $\sim$ ATP consumption) Inset:
  phase diagram showing the IN transition.  }
\label{fig2}       % Give a unique label
\end{figure}

Our model makes several testable predictions. First we find that,
as suggested in Ref.~\cite{hatwalne04}, activity yields a
contribution to the viscosity of an isotropic solution that
diverges at the isotropic-nematic transition (see
Fig.~\ref{fig2}).
Such a diverging viscosity is reminiscent of an equilibrium
liquid-solid transition rather than a liquid-liquid transition,
and is a direct consequence of the active-stresses. Unlike a
liquid-solid transition, however, the divergence is localized at
the IN transition and the viscosity is finite
%% kinematic parameter that is viscosity-like
in the nematic phase. A second novel signature of activity in the
stress relaxation is found in the nematic phase, where the shear
stress acquires a nonequilibrium contribution proportional to ATP
(Adenosine Tri-Phosphate) consumption rate that remains finite for
zero deformation rate. In other words, an active nematic solution
is driven into a state with a non-vanishing macroscopic stress by
the energy input from ATP hydrolysis, even in the absence of an
externally applied mechanical deformation. This non-equilibrium
contribution is also present in the normal stresses in the nematic
phase.

%\paragraph{Microscopic model}

We consider a suspension of polar filaments in a thin film of
height comparable to the length of the filaments (quasi-two
dimensions)  and a constant density  $m$ of motor clusters.  A
concentration $\rho({\bf r},t)$ of filaments is suspended in an
incompressible solvent of viscosity $\eta_0$ characterized by a
fluid velocity ${\bf v}(\bfr,t)$, with $\bnabla \cdot {\bf v} =
0$. Momentum conservation yields
%an equation for the fluid velocity
%
\begin{eqnarray}
\rho_s \left(\partial_t {\bf v} + {\bf v} \cdot \bnabla
{\bf v} \right) &=& \bnabla \cdot \bsigma^s +  \bnabla \cdot
\bsigma^f \;,\;
\label{eqn:momentum}
\end{eqnarray}
with $\rho_s$ the (constant) density of the solution. The solvent
contribution to the stress tensor is
\begin{eqnarray}
\label{fluid_stress} \bnabla \cdot \bsigma^s =  \eta_0 \nabla^2
{\bf v} - \nabla p \nonumber\;, \end{eqnarray}
with $p$ the pressure. The filament contribution, $\bsigma^f$,
must be determined in terms of the driving forces (velocity
gradients, $\kappa_{ij}=\partial_jv_i$, and motor activity,
$\mu~\sim$ ATP consumption rate) and the conserved and order
parameter fields describing the filaments,
\begin{equation}
\label{fils_stress_constitutive} \bsigma^f=\bsigma^f(\rho,{\bf
p},S_{ij};\bm\kappa,\mu)\;,
\end{equation}
where ${\bf p}$ and $S_{ij}$ are the local polarization and
nematic alignment tensor that describe the orientational order of
the filaments. The derivation of this constitutive equation from a
model of filament dynamics is one of the central outcomes of work.

%One important difference as compared to
%passive solutions is that stresses can be induced not just by
%externally applied mechanical deformations (yielding
%$\kappa_{ij}\not=0$), but also by motor activity which maintains
%the system out of equilibrium by supplying energy at a rate $\mu$
%proportional to the ATP consumption rate.

The filaments are modeled as hard rods of {\em fixed} length $l$
and diameter $a$ ($l\gg a$) at position $\bfr$ with filament
polarity characterized by a unit vector $\uvec$. The filament
contribution to the stress tensor is \cite{DoiEdwards}
\begin{eqnarray}
\label{fils_stress_micro}
 \bnabla \cdot \bsigma^f & =& -\int_{\bfr_1}
\int_{\uhat_1} c({\bf r}_1,\uvec_1,t) \Big\langle\delta \left(
{\bf r}-{\bf r}_1-s \uvec \right) {\bf {\cal F}}(s)\Big\rangle_s
\nonumber\;,
\end{eqnarray}
where ${\bf {\cal F}}(s)$ is the hydrodynamic force per unit
length on a rod at position $s$ along the rod, $\langle
..\rangle_s\equiv \int_{-l/2}^{l/2}ds...$, and
$c(\bfr,\uvec,t)$ is the concentration of polar filaments with
position/orientation $\{{\bf r},\uvec\}$. The force on the rod is
specified by its interaction with the solvent, other rods, and the
motor clusters. For low Reynolds numbers $\mbox{Re} \ll 1$,
viscous effects dominate inertia and we can set the left hand side
of Eq.~(\ref{eqn:momentum}) to zero.

%\paragraph{Homogeneous Bulk Steady States.}

We calculate the force per unit length ${\bf {\cal F}}(s)$ by
decomposing a rod into a sequence of beads of diameter $a$ and
solving self-consistently for the flow field around the
rod~\cite{DoiEdwards} on scales much bigger than $a$. The stress
due to the filaments is (to ${\cal O}(\nabla^2)$)
\begin{eqnarray}
\nabla \cdot {\bm \sigma}^f (\bfr,t) &=& \int_{\uvec}  {\rm {\bf
f}}({\bf r},\uvec,t)  \nonumber \\ && -  \int_{\uvec} \big\langle
\big(\frac{s}{l}\big)^2\left(\frac{\uvec \cdot \nabla}{l}
\right){\bm \tau} ({\bf r},\uvec,t)\big\rangle_s\;,
\end{eqnarray}
where
\begin{eqnarray}
\label{Fandtau}
 {\rm {\bf
f}}({\bf r},\uvec,t) &=& c \left[k_B T_a \bnabla \ln c + \bnabla
U_x \nonumber
 - \bfF_{a} ({\bf r},\uvec,t) \right]\;, \\
{\bm \tau} ({\bf r},\uvec,t) &=& c\left[ k_B T_a {\cal R} \ln c +
{\cal R} U_x
- {\bf T}_{a}\right] \times \uvec \nonumber \\
&& - c {\zeta_\perp \over 2} \uvec\uvec (\uvec \cdot \nabla) \cdot
\bfv(\bfr)\;,
\end{eqnarray}
with $U_x(\bfr,\uvec) = k_B T \int_{\uvec'} \int_{\bm \xi} c (\bfr
+ {\bm \xi},\uvec')$  the excluded volume potential.  The force
density has contributions from fluctuations or diffusion (both
thermal and non-thermal - hence the {\em active
  temperature} $T_a \ne T$ ), excluded volume, and motor activity.
  There is
also a viscous contribution to the stress proportional to the
velocity gradient ($\zeta_\perp= 4 \pi \eta_0 l / \ln(l/a)$).

The active force  and torque are  given by
\begin{eqnarray}
&&\bfF_{a}({\bf r},\uvec_1)= - m_0  \int_{\uvec_2} \big\langle\bm
\zeta(\uvec_1) \cdot\bfv_a(1;2) c (\bfr +
\bxi,\uvec_2)\big\rangle_{s_1,s_2}\;, \nonumber\\
\label{FTa} &&{\bf T}_{a}({\bf r},\uvec_1) = - m_0 \int_{\uvec_2}
\big\langle\zeta_r\bomega_a(1;2) c (\bfr + {\bm
\xi},\uvec_2)\big\rangle_{s_1,s_2}\;,
\end{eqnarray}
where $m_0=ma^2$, $(1;2)\equiv(s_1,\uvec_1;s_2,\uvec_2)$,
$\bxi=\uvec_1 s_1 - \uvec_2 s_2$ and ${\bm \zeta}(\uvec) =
\zeta_{\perp}  \left( {\bm \delta} - \uvec\uvec \right) +
\zeta_{\|} \uvec \uvec $, with $\zeta_\perp$, $\zeta_\parallel$
and $\zeta_r$ friction coefficients.
The angular velocity is taken as
\begin{math} {\bomega}_a = 2\left[\gamma_0
+ \left(\uvec_1 \cdot \uvec_2 \right) \gamma_1 \right] \left( \uvec_1 \times \uvec_2 \right)
\end{math}, with $\gamma_0$ and $\gamma_1$ motor-induced rotation rates proportional to ATP consumption,
and tends to align filaments.~\cite{AAMCMTBL} The motor-induced
translational velocity has been derived from a model of motors
walking along the filaments at a mean rate $\beta$. It has the
form $\bfv_a(1;2)={1\over 2} \bfv_{r}+{\bf V}_{m}$,
with~\cite{AAMCMTBL}
\begin{eqnarray}
\label{vel_eq2}
%{\bf v}_{r}&=&\left(\beta-\alpha{\textstyle{\frac{s_1+s_2}
%{2}}}\right)\left(\uvec_2-\uvec_1\right) + \alpha
%{\textstyle\frac{s_1-s_2}{2}}(\uvec_2+\uvec_1)\;, \nonumber\\
{\bf v}_{r}&=&\frac{\tilde\beta}{2}(\uvec_2-\uvec_1)
+\frac{\tilde\alpha}{2l}\bm\xi\;, \nonumber\\
\label{V} {\bf V}_{m} & =& A \left(\uvec_2 + \uvec_1 \right) + B
\left(\uvec_2 - \uvec_1 \right)\;,\nonumber
\end{eqnarray}
where $\tilde\alpha=\alpha(1+\uvec_1\cdot\uvec_2)$ and
$\tilde\beta=\beta(1+\uvec_1\cdot\uvec_2)$. The parameter
$\alpha\sim\beta(a/l)$ is controlled by spatial inhomogeneities
in the motor stepping rate. Momentum conservation yields
expressions for $A$ and $B$. For long thin rods with
$\zeta_\perp=2\zeta_\parallel\equiv 2\zeta$, to leading order in
$\uvec_1 \cdot \uvec_2$, we find $A=-[\beta-\alpha(s_1+s_2)/2]/12$
and $B=\alpha(s_1-s_2)/24$.
%
%\begin{eqnarray}
%A &=& - \left({1 \over 12} \right) \left({\textstyle{\frac{1 - \uvec_1 \cdot \uvec_2}
%{1 - \uvec_1 \cdot \uvec_2/3 }}}\right) \left[\beta-\alpha{\textstyle{\frac{s_1+s_2}{2}}}\right] \\
%B &=&  \left({1\over 12} \right) \left( {\textstyle{1 + \uvec_1
%\cdot \uvec_2 \over 1 + \uvec_1 \cdot \uvec_2/3
%}}\right)\left[\alpha{\textstyle{\frac{s_1-s_2}{2}}}\right]\;,
%\end{eqnarray}
%
When evaluating the contribution to the stress tensor, only terms
up to first order in $\uvec_1\cdot\uvec_2$ are retained in the
active force $\bm\zeta(\uvec_1)\cdot{\bf v}_a(1;2)$ exerted by a
motor cluster on the filament in the first of Eqs.~(\ref{FTa}).
This approximation only affects the numerical values of the
coefficients in the stress tensor, not its general form.
%The corresponding
%terms bundle parallel filaments, while the $\beta$ terms separate
%antiparallel filaments.

% In the following all lengths are measured in units of the
%filament length, $l$.
The concentration $c(\bfr,\uvec,t)$ of polar filaments  satisfies
a local conservation law,
\begin{equation}
\partial_t c + \bnabla \cdot ({\bf v} c) + {\cal R} \cdot \left( \bomega c \right)+ \bnabla \cdot {\bf J}
+ {\cal R} \cdot {\cal J} =0\;, \label{smoluchowski}
\end{equation}
where $\bomega=\uvec \times {\bm \kappa} \cdot \uvec$ and ${\cal
R}=\uvec\times\frac{\partial}{\partial\uvec}$. The translational
and rotational currents in Eq.~(\ref{smoluchowski}) contain
diffusive, excluded volume and active contributions,
\cite{AAMCMTBL}
\begin{math}
{\bf J} = -\bm\zeta^{-1}(\uvec)\cdot{\rm {\bf
f}}(\bfr,\uvec)\;, \;
{\cal J} = -\zeta_r^{-1}\bm\tau(\bfr,\uvec)\;,
\end{math}
with ${\rm {\bf f}}$ and $\bm\tau$ given by Eqs.~(\ref{Fandtau}).

The conserved and broken symmetry fields are the  $\mbox{density}
\; \rho(\bfr,t)$, $\mbox{polarization} \; {\bf p}(\bfr,t)$, and
$\mbox{nematic order} \; S_{ij}(\bfr,t)$,  defined as moments of
the probability distribution,
\begin{math}
\int_\uvec c(\bfr,\uvec,t)=\rho(\bfr,t)\; ,
\end{math}
\begin{math}
\int_\uvec \, \uvec \, c(\bfr,\uvec,t)=\rho(\bfr,t){\bf
p}(\bfr,t)\;,
\end{math}
\begin{math}
\int_\uvec \, \hat{Q}_{ij} \,
c(\bfr,\uvec,t)=\rho(\bfr,t)S_{ij}(\bfr,t)\; \, ,
\label{orient_moments} \nonumber
\end{math}
where $\hat{Q}_{ij}=\uhat_i\uhat_j-\frac{1}{2}\delta_{ij}$.
Continuum equations for these fields are obtained from
Eq.~(\ref{smoluchowski}) by the coarse-graining procedure
described in Ref.~\cite{AAMCMTBL}.

The constitutive equation for the stress tensor $\bm\sigma^f$ is
obtained by evaluating the right hand side of
Eq.~(\ref{fils_stress_micro}). For simplicity we consider
spatially homogeneous solutions in the presence of a constant
velocity gradient $\kappa_{ij}$. To lowest order in gradients, the
three contributions to the stress tensor of the filaments are
%(1-active forces, 2-passive torques and 3-passive forces)
%
\begin{equation}
\sigma_{ij}^f (\bfr,t) = \sigma_{ij}^P (\bfr,t) + \sigma_{ij}^A (\bfr,t) + \sigma_{ij}^v (\bfr,t)\; ,
\end{equation}
with passive and active contributions, $\bsigma^P$ and
$\bsigma^A$, given by
\begin{eqnarray}
\sigma_{ij}^P & =& 2 k_B T_a \rho\Big[\Big(1 - {\rho \over
\rho_{IN}} \Big)  S_{ij}
-{ \rho \over 2\rho_{IP}} \big( p_i p_j - {1 \over 2} p^2 \delta_{ij} \big)\nonumber \\
&&+\frac{1}{2}\delta_{ij}\Big(1+\frac{l^2\rho}{\pi}\big(1-\frac{2}{3}S^2\big)\Big)\Big]\;,  \label{equil_stress}\\
\sigma_{ij}^A &=&  \mu k_B T_a \rho^2\Big[{8  \over 9}
\big(S_{ij}+\frac{1}{2}\delta_{ij}\big) + {2 \over 3}   \big( p_i
p_j + {1 \over 2} p^2 \delta_{ij} \big) \nonumber \\ && +
\frac{1}{9}(1-S^2)\delta_{ij} \Big]\;, \label{act_stress}
\end{eqnarray}
where $\mu = {{m_0}
  \alpha l^3 \over 48 D}$, $\rho_{IP}=D_r/(m_0\gamma_0l^2)$, and
$\rho_{IN}=\rho_N/[1+\rho_N l^2m_0\gamma_1/(4D_r)]$ is the
density for the isotropic-nematic (IN) transition at finite motor
density \cite{AAMCMTBL}, with $\rho_N=3\pi/(2 l^2)$ the density
of the IN transition in passive systems.
  Finally, the viscous contribution to the
stress is
\begin{eqnarray}
\sigma_{ij}^v &=& {l \rho \zeta_\perp \over 24} \Big[{1 \over 2}
\big(\kappa^s_{ij} +{\textstyle{1\over 2}}\kappa_{kk}\delta_{ij}\big)  +  {1
\over 3} \big( \delta_{ij} \kappa_{kq} S_{qk} + S_{ij}
\kappa_{kk} \big) \nonumber \\
&& +  \kappa^s_{ik}S_{kj} + \kappa^s_{jk}S_{ki}
\Big]\;,\label{visc_stress}
\end{eqnarray}
with $\kappa^s_{ij}=(\kappa_{ij}+\kappa_{ji})/2$.

In agreement with Ref.~\cite{hatwalne04}, we find that active
units generate force dipoles in the fluid yielding contributions
to the stress which are equilibrium-like, i.e., have the same form
as those appearing in an equilibrium solution, but with new
contractile stresses ($\alpha>0$) which have no analogue in their
equilibrium counterparts.

%As pointed out in Ref.~\cite{hatwalne04},
%active units (e.g., a pair of filaments coupled by a motor cluster)
%act as force dipoles in the fluid, yielding an active contribution to
%the stress which is equilibrium-like in the sense that it contains
%terms of the same form as those appearing in the stress of an
%equilibrium solution of passive rods. Motor-filament units yield
%contractile stresses ($\alpha>0$), opposing the relaxation of an
%applied shear and increasing the shear viscosity of the fluid.

For a homogeneous solution $\rho={\rm constant}$ and the equations
for polarization and nematic order parameter are obtained by
averaging Eq.~(\ref{smoluchowski}) over $\uvec$,
%($\rho={\rm constant}$)
%
\begin{eqnarray}
\partial_tp_i&=&-\Omega_i-D_r\Big(1-\frac{\rho}{\rho_{IP}}\Big) p_i
\nonumber \\ && +2 D_r \Big( { 2 \rho \over \rho_{IN}} - {\rho
\over \rho_{IP}}\Big)
S_{ij}p_j\;, \label{eq:pdyn}  \\
\partial_t
S_{ij}&=&-\Omega_{ij}-4D_r\Big(1-\frac{\rho}{\rho_{IN}}\Big)
S_{ij} \nonumber \\ && + { 2 D_r\rho \over \rho_{IP}}
\Big(p_ip_j-\frac{1}{2}\delta_{ij}p^2\Big)\;. \label{eq:Sdyn}
 \end{eqnarray}
with
\begin{math}
\rho\Omega_{i}=\int d\uvec \uhat_i\bm{\mathcal R}\cdot(\bomega f)
\end{math} and \begin{math}\rho\Omega_{ij}=\int d\uvec \hat{Q}_{ij}\bm{\mathcal
R}\cdot(\bomega f)\;.
\end{math}
We find
\begin{eqnarray}
&&\Omega_{i}=-\kappa_{ij}p_j+\frac{1}{2}\Big[\kappa^s_{ij}p_j+{\textstyle\frac{1}{2}}\kappa_{kk}p_i\Big] \;,\\
&&\Omega_{ij}=-\frac{1}{2}\big[\kappa^s_{ij}-{\textstyle\frac{1}{2}}\delta_{ij}\kappa_{kk}\big]
-\big(\kappa_{ik}S_{kj}+\kappa_{jk}S_{ki}\big)\nonumber\\
&& +\frac{1}{3}\Big[S_{ij}\kappa_{kk}+\delta_{ij}S_{kl}\kappa_{kl}
+2(S_{ik}\kappa^s_{kj}+S_{jk}\kappa^s_{ki})\Big]. \label{Omega_ij}
\end{eqnarray}

The homogeneous (bulk) steady-states are obtained by setting the
rhs of the order parameter equations to zero. We find (I)sotropic
($\bfp=0,S_{ij}=0$), (N)ematic ($\bfp=0,S_{ij} \ne 0$) and
(P)olarized ($\bfp \ne 0,S_{ij} \ne 0$) phases. It can be shown
using Eqs. (\ref{equil_stress},\ref{act_stress}) that in a {\em
  stationary} bulk fluid the passive contribution to the stress tensor
  is identically zero
in each phase, while the active contribution is non-zero.  In the
following we consider the geometry of pure shear flow with
$\kappa_{ij}=\dot{\epsilon}\delta_{ix}\delta_{jy}$
% where
%
%\begin{eqnarray}
%&&\Omega_x=-\frac{3}{4}\dot{\epsilon}p_y\;,\hspace{0.3in}
%\Omega_y=\frac{1}{4}\dot{\epsilon}p_x\;,\\
%&&\Omega_{xy}=\Omega_{yx}=-\frac{1}{4}\dot{\epsilon}-\dot{\epsilon}S_{yy}\;,\\
%&&\Omega_{xx}=-\Omega_{yy}=-\dot{\epsilon}S_{xy}\;.
%\end{eqnarray}
%
and discuss the {\em linear} viscoelastic response of the active
solution in the isotropic (I), polarized (P) and nematic (N)
phases.

%\begin{widetext}
%

\paragraph{Isotropic phase}

In the isotropic phase close to the IN transition a shear flow
builds up nematic order, yielding a non-zero value for $S_{ij}$ to
${\cal O} (\dot{\epsilon})$.  The shear stress is linear in the
strain rate, $\sigma_{xy} = {\cal O} (\dot{\epsilon})$, while the
normal stress is quadratic, $\sigma_{xx}-\sigma_{yy}= {\cal O}
\left(\dot{\epsilon}^2\right)$. To linear order we obtain
$\big(\partial_t +1/\tau_A\big)S_{xy}=\dot\epsilon/4$, with
\beqa \frac{1}{\tau_A} =\frac{4}{\tau_0}
\Big(1-\frac{\rho}{\rho_{IN}}\Big)\;, \eeqa
where $\tau_0=1/D_r$ and $D_r=k_BT_a/\zeta_r$ the rotational
diffusion constant. The time scale $\tau_A$ diverges as we
approach the {\em active} IN transition. In a sheared sample, the
total stress (filaments + solution) is
\begin{equation}
\sigma_{xy}=\tilde{\eta}\dot{\epsilon}+2k_BT\left(1-{ \rho \over
\rho_{IN}}\right)\rho S_{xy}+{ 8 \over 9 }\mu k_B T \rho^2
S_{xy}\;,
\end{equation}
where $\displaystyle\tilde{\eta} = \eta_0 ( 1 + {\pi l^2 \rho
\over 24})$. The first contribution is from the solvent and the
viscous stress, the second is from the passive stress, and the
third from active stresses.

%The frequency-dependent shear
%viscosity of the active solution, defined via $ \sigma_{xy} =
%\eta_A (\omega) \dot{\epsilon}$ is given by
%
%\beq \eta_A(\omega)=\tilde{\eta} +\frac{k_B T_a\rho/(8
% D_r)}{1+i\omega\tau_A}\Big[1+\frac{16}{9}\tau_A D_r\mu\Big]\;.
%\eeq
%
%Taking the limit $\omega \rightarrow 0$,

For an oscillatory applied shear we define the frequency dependent
shear viscosity $ \sigma_{xy}(\omega) = \eta_A (\omega)
\dot{\epsilon}$, with low frequency limit
\beq \eta_A=\tilde{\eta}+\frac{k_B T\rho}{8 D_r }\Big(1
+\frac{16}{9}\tau_A D_r\mu\Big)\;. \eeq
This {\em diverges} as we approach the $I-N$ transition (see
Fig.~\ref{fig2}). This behavior should be contrasted to that of a
passive solution of rods near the \emph{equilibrium} I-N
transition. In this case the stress relaxation time
%whose frequency-dependent viscosity is given by
%
%\beq \eta_P(\omega)=\tilde{\eta}+\frac{k_BT\rho/( 8 D_r)}{1+i\omega\tau_P}\;,
%\eeq
%with
$\tau_P=D_r^{-1}/(1-\rho/\rho_N)$ also diverges at the IN
transition, but the zero-frequency viscosity,
$\eta=\tilde{\eta}+k_B T\rho/(8 D_r)$, remains {\em finite}, as
required by the Fluctuation-Dissipation Theorem.

\paragraph{Nematic phase}

In the nematic phase, there is the possibility of alignment of the
nematic director by the shear flow. The director ${\bf n}$ and the
magnitude $S$ of the nematic order parameter are defined by
$S_{ij} = S \left(\nhat_i\nhat_j -
 \delta_{ij}/2 \right)$.  The equation of
motion for the director in a steady shear flow is obtained from
Eq. ~(\ref{eq:Sdyn}),
\begin{equation}
\partial_t \nhat_i = \left(\delta_{ij}-\nhat_i \nhat_j \right) \nhat_k {\Omega_{jk} \over
S}\;.
\end{equation}
Defining $\nvec=(\cos \theta, \sin \theta)$, we obtain a steady state
solution for the director orientation given by $ \cos 2\theta = 2 S$.
For $S > 1/2$ there are no steady-state solutions, possibly pointing
to the existence of periodic or chaotic solutions characterized by
``wagging'' or ``kayaking" of the nematic director \cite{briels_05}.

The steady-state expression for the stress
tensor in the {\em active} nematic state is obtained from
Eqs.~(\ref{equil_stress}) and (\ref{act_stress}) as
\begin{equation}
\sigma_{ij} = -{ k_B T  \rho \over 2 D_r } \left( {\Omega_{ij}
\over \rho} \right) + \sigma_{ij}^A + \sigma_{ij}^v \; ,
\end{equation}
from which  the 6 Leslie coefficients of the active nematic can be
obtained using Eqs.~(\ref{act_stress}),(\ref{visc_stress}) and
(\ref{Omega_ij})~\cite{DoiEdwards,TBLMCM07}.

The novel nature of the constitutive equation is best illustrated
by the simple case of an active nematic in the {\em flow-aligning}
regime ($S<1/2$), where the steady-state shear stress is given by
\begin{equation}
\sigma_{xy} = \eta \dot{\epsilon} + \chi \mu\;,
\label{new_constit}\end{equation}
where \begin{math} \eta = \tilde{\eta} + {k_B T \over 2 D_r} \rho
  \left( {1 \over 4} - S^2 \right) \left(1 + D_r \tau'_A { 16
      \tilde{\alpha} \rho \over 9 } \right)\; \mbox{and} \; \chi = {8
    \over 9} k_B T \rho^2 {S } \sqrt{{1 \over 4} - S^2}\end{math},
with $1/\tau'_A = 8 D_r \left( \rho/\rho_{IN} -1 \right)$.  The
magnitude of  nematic order $S$ relaxes on a time
scale $\tau'_A$ while the director relaxes on the shear time
scale $\dot{\epsilon}$ leading to non-monotonic stress relaxation.

The signature of this constitutive equation is a shear stress that
does not vanish for zero deformation rate.  This is because the active
filament system is being driven out of equilibrium by two sources of
energy.  One external, due to the shear, and the other internal, due
to the activity of the motors (see Fig.~\ref{fig2}).  The viscosity
diverges as the IN transition is approached from the nematic side, but
it decreases dramatically as one goes deeper into the nematic phase
(see Fig.~\ref{fig2}). This is a direct consequence of flow alignment
as $S$ increases with density. All corrections to $\eta$ vanish at
$S=1/2$, where the flow-aligned regime ceases to exist.

Another signature of a nematic phase is that the normal stresses
are of first order in the shear rate. In addition, in an active
nematic one obtains an anomalous constitutive equation,
\begin{equation}
\sigma_{yy}-\sigma_{xx} = \eta_N \dot{\epsilon} + \chi_N \mu\;,
\end{equation}
where $\eta_N = {k_B T  \over 2 D_r} \rho {S } \sqrt{{1 \over 4} -
S^2} $ and \begin{math} \chi_N = {8\over 9} k_B T \rho^2 2 S^2
\end{math}.

Both the anomalous stresses (shear and normal) should be easily
observed by performing linear rheological experiments on active
filaments in the nematic phase at varying shear rates.
%Recent mechanical experiments of supported cell cultures also seem to
%observe such active stresses in vivo~\cite{Amblard06}.

\paragraph{Polarized phase}
In a polarized state, with ${\bf p}=p_0{\bf \hat{p}}$, a uniform
density and a constant velocity gradient, the unit vector ${\bf
\hat{p}}$ satisfies the dynamical equation
%satisfies $\bm\nabla\cdot{\bf \hat{p}}$. Its dynamics is
%governed by the equation
%
\begin{equation}
\partial_t\hat{p}_i+\omega_{ij}\hat{p}_j=\lambda\delta^T_{ij}\kappa^s_{jk}\hat{p}_k\;,
\end{equation}
where $\lambda=1/2$, $\omega_{ij}=(\kappa_{ji}-\kappa_{ij})/2$ and
$\delta^T_{ij}=\left(\delta_{ij}-\hat{p}_i\hat{p}_j\right)$.
Voituriez et al. ~\cite{Voituriez} have recently suggested that
spontaneous flow can be obtained in active polar materials,
corresponding to a state with finite velocity gradients and flow
alignment. Flow alignment, however, can only occur for
$\lambda>1$, while the present (low density) calculation yields
$\lambda=1/2$.  Corrections to $\lambda$ can be obtained by
incorporating the fact that fact that the alignment tensor is
slaved to the polarization field.
%In the polarized state the alignment tensor is slaved to the polarization
%field. This effects yields correction to the parameter $\lambda$ that
%becomes a function of $S$ and therefore $p_0$.
However an analysis of such corrections shows that they fail to
increase $\lambda$ to values larger than 1,
%Therefore for the (low density) microscopic description presented here,
%we have not been able
%to obtain a steady flowing uniformly polarized state.
suggesting that no steady uniformly flowing polarized state 
exists for a thin film 
%in two dimensions 
in the low density limit presented here.  An
interesting alternative is the possibility of periodic or chaotic
states~\cite{rheochaos}.

In summary, we have used a molecular model to study the macroscopic
mechanical response of active filament solutions in both isotropic and
ordered states.  Motor activity leads to a novel coupling of
mechanical properties to order and to anomalous constitutive equations
in the liquid crystalline state. The theoretical framework developed
here can be generalized to consider stress inhomogeneities. This is
relevant for understanding the microrheology of active filament
systems where new behavior is expected even in the isotropic
regime~\cite{TBLMCM07}.

\paragraph*{Acknowledgments}
TBL acknowledges the support of the Royal Society and the National
Science Foundation under Grant No. PHY99-07949. MCM acknowledges
support from the National Science Foundation, grants DMR-0305407 and
DMR-0219292.

\end{document}